\documentclass{aastex}
\usepackage{emulateapj5}                                                        

\newcommand{\hh}    {HH~80N}
\newcommand{\kms}   {km~s$^{-1}$}

\newcommand{\jy}    {Jy~beam$^{-1}$}

\newcommand{\yr}    {yr$^{-1}$}
\newcommand{\vlsr}  {$v_{\rm LSR}$}
\newcommand{\lo}    {$L_{\sun}$}
\newcommand{\mo}    {$M_{\sun}$}

\newcommand{\et}    {et al.}
\newcommand{\eg}    {e.\,g.,}
\newcommand{\ie}    {i.\,e.,}

\newcommand{\rod}   {Rodr\'{\i}guez}

\newcommand{\hco}   {HCO$^+$}

\newcommand{\J}[2]  {\mbox{$J$=#1--#2}}

\begin{document}

\received{\rule{3cm}{0.5pt}}
\accepted{\rule{3cm}{0.5pt}}

\lefthead{Girart \et}
\righthead{
Star Formation Signatures in the Condensation Downstream of HH~80N
}

\slugcomment{
{\em The Astrophysical Journal Letters, accepted},
Printed \today
}

\title{
Star Formation Signatures in the Condensation Downstream of HH~80N
}

\author{
J. M.        Girart\altaffilmark{1,2} 
R.        Estalella\altaffilmark{1}, 
S.             Viti\altaffilmark{3},
D. A.      Williams\altaffilmark{3}, and
P. T. P.         Ho\altaffilmark{4}
}

\altaffiltext{1}{
Departament d'Astronomia i Meteorologia, Universitat de Barcelona, 
Av.\ Diagonal 647, 08028 Barcelona, Catalunya, Spain; jgirart@am.ub.es
}
\altaffiltext{2}{
Department of Astronomy, University of Illinois, 1002 W. Green St, Urbana, 
IL 61801, USA}
\altaffiltext{3}{
Department of Physics and Astronomy, University College London,
London, WC1E 6BT, England}
\altaffiltext{4}{
Harvard-Smithsonian Center for Astrophysics, 60 Garden Street, Cambridge, 
MA 02138, USA
}

%\clearpage

\begin{abstract}

\hh\ is one of the Herbig-Haro objects that have associated quiescent 
dense clumps.  We report CO and CS BIMA observations that reveal star formation 
within the \hh\ dense clump. The CO emission reveals clearly a bipolar
molecular outflow centered on the dense clump.  The CS emission traces a 
ring-like structure of radius $\simeq 0.24$~pc.  The CS kinematics shows that
the ring is collapsing with an infall speed of $\sim 0.6$~\kms.  The required
mass to produce the collapse is in agreement with previous ammonia observations
of the 20~\mo\ core, which is embedded within the CS structure.  However, we 
cannot discard that the ring structure is expanding driven by protostellar 
winds, if the CS abundance if unusually high and the CO momentum rate is much 
higher than that measured, due to inclination and optical depth effects.  The 
properties of the molecular outflow and of the dense core suggest that it 
harbors a Class 0 object.  There are also signatures of interaction of the 
HH~80/81/80N outflow with the dense gas. In particular it is possible that the 
HH~80/81/80N outflow has triggered or at least speed up the star formation in
this region.

\end{abstract}


%\clearpage

\section{Introduction\label{intro}}

HH~80/81/80N is a spectacular and very luminous HH outflow spanning about
5~pc and exhibiting proper motions, close to the exciting source, of up to
1400~\kms\ \citep{reipurth88, marti93, marti95, marti98, heathcote98,
molinari01}.  \hh, the northern counterpart of HH~80/81, is optically
invisible, due to the high extinction at this position (Mart\'{\i} \et\ 1993),
but its radio and far-infrared properties confirm its HH nature (Mart\'{\i} 
\et\ 1993; Molinari \et\ 2001).  Far-infrared ISO observations show that the
strong FUV field radiated by the HH~80/81/80N outflow have induced the
formation of a PDR (photo-dissociation region) in its immediately
surrounding region (Molinari \et\ 2001).

Molecular line observations towards \hh\ \citep{girart94, girart98} showed 
that it belongs to the class of Herbig-Haro (HH hereafter) objects, typically 
of high excitation, that are found to have starless quiescent dense clumps 
ahead of them. The strong radiation arising 
from the HH objects penetrates the clumps, releasing the icy mantles from 
the dust grains and inducing a photon-dominated chemistry, which causes 
significantly altered chemistry \citep{taylor96, viti99}.  Recent observations 
towards HH~2 have confirmed this chemical scenario (Girart \et\ in 
preparation).  However, the presence of IRAS~18163-2042 associated with the 
\hh\ dense clump suggested that it could be tracing an embedded star 
(Girart \et\ 1998).

This paper presents CS and CO BIMA observations that are part of a molecular 
line survey towards \hh\ whose complete results will be reported in a 
forthcoming paper.  The observations presented in this paper show infall and 
outflow signatures associated with the dense clump downstream of \hh.  The 
high angular resolution provided by millimeter interferometers, as in this 
paper, have allowed astronomers to resolve spatially infall kinematical 
signatures in some low mass star forming cores \citep{zhou96, ohashi97, 
saito96, momose98, hoger01}.  Typically, the infalling gas arises from scales 
of $\sim1000$--$3000$~AU and the mass infall rates are 
1--10$\times10^{-6}$~\mo~yr$^{-1}$. Interestingly, the infall signatures found 
in \hh\ are similar to that from the dense cores in L1544 \citep{ohashi99} and 
HH~1/2 \citep{torrelles94}, which show infall signatures in the form of an 
almost ring-like structure in the position-velocity diagram on even larger 
scales, $1.5\times10^4$ and $4.5\times10^4$~AU, respectively.

\section{Observations and Results\label{resul}}

The observations were carried out between 1999 May and September with the
10-antenna BIMA array\footnote{The BIMA array is operated by the
Berkeley-Illinois-Maryland Association with support from the National
Science Foundation.} at the Hat Creek Radio Observatory.  The phase
calibrator used was QSO 1733-130.  Absolute flux calibration was performed 
by observing Uranus and Mars.  The phase center of the observations was
$\alpha (J2000) = 18^{\rm h}19^{\rm m}18\fs62$;  $\delta (J2000) =
-20\arcdeg40'55''$.  Two different frequency setups were used in order to
observe the CS \J{2}{1}\ and CO \J{1}{0}\ lines, which provided a spectral
resolution of 0.3 and 0.5~\kms, respectively.  Maps were made with the 
$(u,v)$ data weighted by the associated system temperatures.

\subsection{The CS \J{2}{1}\ Emission}

The CS emission arises from an elongated structure in the NW to SE
direction ($PA\sim 125\arcdeg$), consisting of two well differentiated
structures morphologically and kinematically (see Figures~\ref{fmapa},
\ref{fposvel}):  (1)  The main core is centered at the position of the
ammonia core \citep{girart94}.  The position-velocity plot along the major 
axis of the CS emission shows an ring-like structure, roughly symmetrical about
$V_{LSR}\simeq11.4$~km~s$^{-1}$. This structure can be explained by inward
(or outwards) motions in the condensation (see \S~\ref{cau}) and resembles
that from the starless core L1544 \citep{ohashi99}. The physical scale of
the dense main core traced by the CS emission, $\sim 0.48\times0.17$~pc
(assuming a distance of 1.7~kpc: Mart\'{\i} \et\ 1993) is larger than that
from the NH$_3$ (1,1) \citep{girart94} and the \hco\ \J{3}{2}\
(Girart \et\ 1998).  This could be due to the higher critical density of
these two lines with respect to the CS \J{2}{1}\ transition.  (2) A small
redshifted ''arm'' (hereafter RSE arm) is located SE of the main core and 
$\sim 15''$ north of the 6~cm continuum emission.  The RSE arm exhibits a 
clear linear velocity gradient with increasing velocity towards the east, 
\ie\ as it gets closer to \hh. Table~\ref{fisica} shows the physical 
parameters of the core traced by the CS \J{2}{1}\ line.  Of the total mass 
measured, 12~\mo\ (assuming $X({\rm CS})=2\times10^{-9}$), around 10~\mo\ 
belong to the main core, and 2~\mo\ to the RSE arm.

\subsection{The molecular outflow}

The blueshifted and redshifted CO emission traces a clearly bipolar outflow
along the E-W direction (lobes BE and RW) and approximately centered at 
the peak intensity of the CS integrated emission (Fig.~\ref{fmapa}).  The
presence of molecular outflows is a clear signpost of star formation within 
the dense core.  The outflow axis is not perpendicular to the major axis of 
the main core.  However, our angular resolution traces scales 
($\sim2\times10^4$~AU) much larger than the scales at which the collimation
process occurs, $\sim 1$~AU \citep{shu95}.  Table~\ref{fisica}\ shows the 
physical parameters of the molecular outflow.  There is marginal evidence of 
another bipolar outflow in the N-S direction and also centered in the 
dense core.  In addition to the molecular outflow, there is a redshifted clump 
that coincides not only spatially with the RSE arm but also kinematically, 
suggesting that this CO emission traces the same gas as the CS at this 
location.

\section{Discussion}

\subsection{The collapse of the main core\label{cau}}

The position-velocity plot along the major and the minor axes of the CS 
emission (Fig.~\ref{fposvel}) clearly suggests that the gas is either 
collapsing or expanding radially and that the emission arises from a ring-like 
or toroidal structure.  A similar structure has being observed towards the 
starless core L1554 from CCS observations \citep{ohashi99}.  The ratio between 
the major and minor axes of the CS emission gives an inclination angle of 
$\ga 70\arcdeg$.  As pointed out by \citet{ohashi99}, the fact that the 
emission is detected towards a ring-like structure is probably due to a 
chemical effect, \ie\ a decrease of the CS abundance toward the center, 
possibly because of gas depletion onto grains at higher densities
\citep{rawlings92}.

We considered a model of a spatially thin ring with both infall and
rotation, similar to that of \citet{ohashi97, ohashi99}, seen edge-on by
the observer.  The parameters of the model were the inner and outer radii
of the ring, $R_{\rm inn}$ and $R_{\rm out}$, the infall and rotation
velocities, $V_{\rm infall}$ and $V_{\rm rot}$, and the intrinsic line width 
$\Delta V_{\rm core}$.  The position-velocity diagram was computed along the
projected major and minor axes of the ring, assuming that the intensity was 
constant for $R_{\rm inn}<R<R_{\rm out}$, to compare with the observed 
kinematics.  The best fit was obtained for velocities $V_{\rm
infall}=0.59$~km~s$^{-1}$, $V_{\rm rot}\lesssim0.20$~km~s$^{-1}$, and
radii of the ring $R_{\rm inn}=0.19$~pc, $R_{\rm out}=0.29$~pc (see
Table~\ref{fisica}). The observed and synthetic position-velocity diagrams
are shown in Fig.~\ref{fposvel}. This model also applies if the ring is 
expanding instead of contracting. In this case, the derived infall velocity 
should be taken as the expantion velocity.

Can the protostellar wind cause the expansion of a ring-like structure?  
The momentum rate of the ring (see Table~\ref{fisica}) is
$\sim 2\times10^{-5}$~\mo~\kms~yr$^{-1}$.  On the other hand, the momentum
rate of the protostellar wind, assuming that it is the same as that from
the molecular outflow, is $\sim 1\times10^{-5}$~\mo~\kms~yr$^{-1}$.  
Although both values are similar, only a small fraction of the wind momentum
rate would be transfered to the dense core.  First, the dense core, as traced 
by the CS, shows a relatively flattened structure.  Assuming that the ring 
structure traced by the CS is a torus, then the inner and outer radius derived 
from the model will imply a semi-aperture angle of the torus as seen from 
the center of $12\arcdeg$.  Second, the momentum rate distribution of the wind
is expected to be anisotropic, \ie\ most of the momentum rate of the wind is 
carried out along the axis of the outflow, \eg\ \citet{shu00}.  
The X-wind model \citep{shu95} predicts that at a distance of 0.24~pc the
fraction of wind mass moving at equatorial angles lower than $12\arcdeg$ is 
about 2\% of the total wind mass.  We expect the momentum rate to scale 
similarly.  Thus the wind that could interact with the dense core has roughly 
a momentum rate $2 \times 10^{-7}$~\mo~\kms~yr$^{-1}$, which is much lower 
than that measured from the CS kinematics, assuming a typical CS abundance of 
$X({\rm CS})=2\times10^{-9}$.  Thus, the wind would be able to cause the 
expansion of the ring only if the CS has an unusually high abundance, and if 
the true momentum rate is much higher than that measured, because of the 
outflow inclination and the optical depth. 

Can the ring-like structure be in contraction? \citet{girart94} found that 
the ammonia emission arises from a smaller region than the CS \J{2}{1}\ and 
traces a 20~\mo\ core (see \S~\ref{resul}).  This mass is similar to that
required to cause free-fall collapse for the infall velocity measured at a 
radius of $\sim 0.24$~pc.  However, the gas pressure and magneticc field should
be taken into account. For the intrinsic line width obtained from the model, 
the virial mass without magnetic field for the CS main core is 9~\mo.  This
indicates that in the absence of the magnetic field the gas pressure cannot
prevent the collapse.  Assuming that the ring-like structure is contracting, 
we can provide a rough upper limit on $|{\bf B}|$, \ie\ the main core should 
be supercritical and the total virial mass (including the magnetic fields) 
should be no more than the free-fall mass.  Following the formulae given by 
\citet{crutcher99} and \citet{mckee93} we find that for a mass of 20~\mo\ the 
collapse is possible for $|{\bf B}| < 50 \mu$G.  Magnetic fields of 
$\la50\mu$G have been measured in several molecular clouds with similar sizes 
and line widths as those of \hh\ \citep{crutcher99}.  
Thus, it seems that there is enough mass to cause collapse while, on the other
hand, an expansion would require an unusually high CS abundace and a high
correction to the observed CO momentum rate.  Therefore, we favour that the CS 
kinematics is tracing infall.

Compared with most of the low mass regions where a kinematical infall
signature has been found (see \S~\ref{intro}), the \hh\ main core has a
larger mass infall rate (see Table~\ref{fisica}).  This is probably because 
it is also a more massive region.  Another example is the infalling NH$_3$ 
core in HH~1/2 \citep{torrelles94}, where the infall material also comes from 
a ring-like structure (toroid) with a similar size and mass as the \hh\ main 
core.  Interestingly, the \hh\ and HH~1/2 star forming cores show clear 
differences with the other collapsing ring-like structures detected so far.  
Here, the infalling motions are clearly supersonic with no rotation observed, 
whereas the infall motions in the starless core L1544 are subsonic and of the 
order of the rotation speeds \citep{tafalla98, ohashi99, williams99}.  These 
differences could be due to an evolutionary effect (L1544 is still a starless 
core) and/or to the different masses involved in the collapse.  The powering
source of HH~1/2 is a protostar with an far-IR luminosity of $\sim 50$~\lo
\citep{harvey86}, but what about \hh?

\subsection{The evolutionary stage of the \hh\ main core}

IRAS~18163-2042 has a luminosity of $\sim 20$~\lo\ \citep{girart94}, but 
because of the large position uncertainty it is possible that both the PDR 
region heated by \hh\ (Molinari \et\ 2001) and the protostar embedded in the 
infalling core contribute to this luminosity, Therefore, this value is a 
rough upper limit of the protostellar luminosity.

A detailed study of the outflow activity, from a homogeneous sample of
low-luminosity embedded YSOs, done by \citet{bontemps96} found that the 
relation between the outflow momentum flux, the envelope mass and the 
bolometric luminosity depends on the evolutionary stage of the protostars.  
In particular, the relation between $F_{\rm CO} \, c/L_{\rm bol}$ and 
$M_{\rm env}/L_{\rm bol}^{0.6}$, which should be almost
free of any distance and luminosity effects, is useful to distinguish the
evolutionary stage of the low mass YSOs, that is, whether they are in the
Class 0 or I stage.  The values of the measured envelope mass and the corrected
momentum rate, $F_{\rm CO}=10 \dot{P}$ (the same inclination and opacity  
correction factor used by Bontemps \et\ 1996), suggest that, for the possible 
range of bolometric luminosities ($\la$20~\lo), the \hh\ dense core is in the 
Class 0 evolutionary stage.  \citet{bontemps96} also found that the momentum 
flux and the envelope mass are well correlated for both Class 0 and I objects.  
However, taking into account the momentum flux of the \hh\ molecular outflow, 
the dense core is an order of magnitude more massive than what is expected 
from this correlation.  The excess of envelope mass with respect to the 
outflow strength could be explained if there is multiple star formation going 
on (\eg\ if at unresolved scales the core splits into several clumps, with some 
of them still pre-stellar).

\subsection{Interaction between \hh\ and the molecular cloud}

\begin{figure} 
\epsscale{0.5}
\plotone{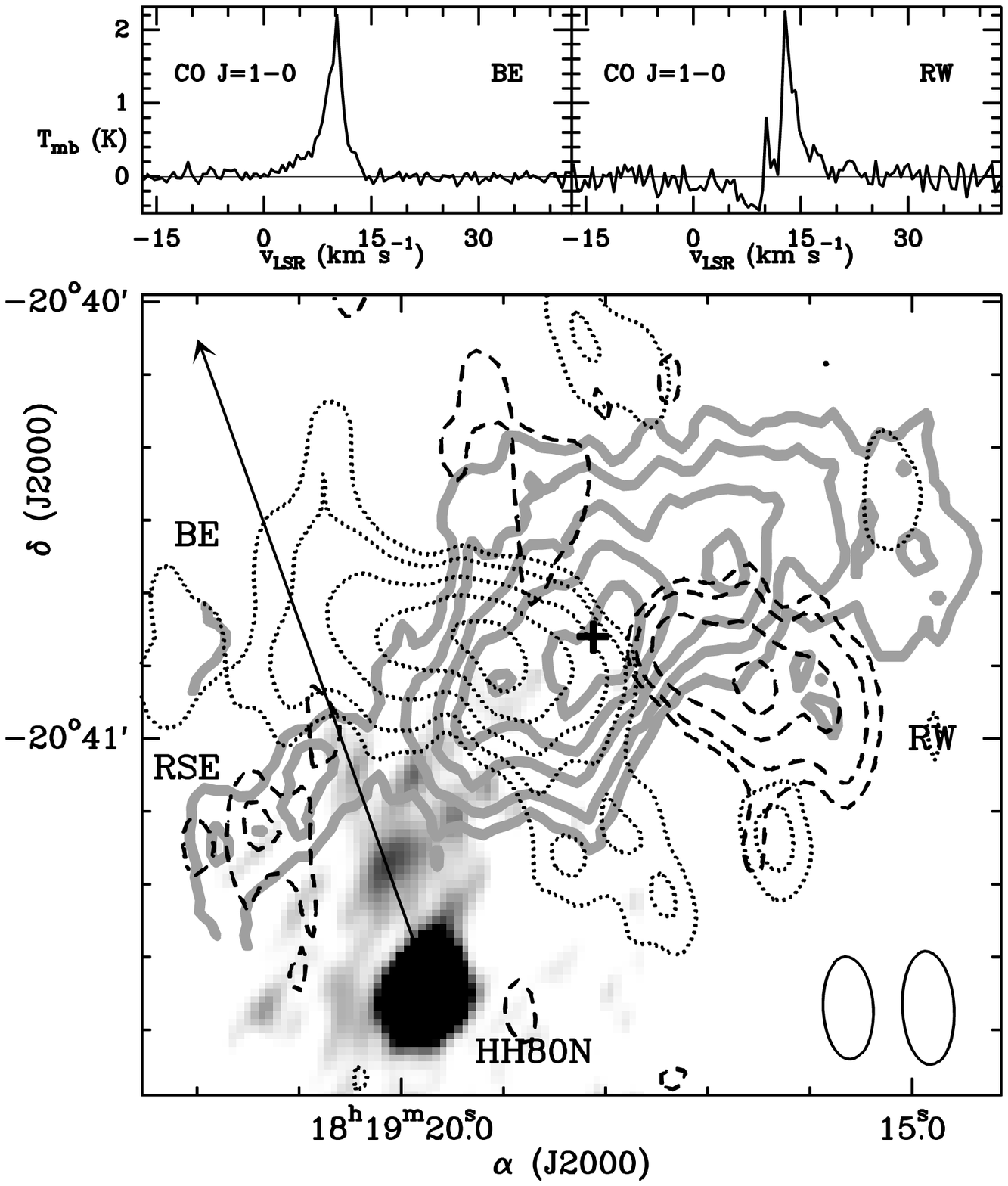} 
\figcaption[fmapa.ps]
{Superposition of the: 
{\it gray scale} map of the VLA 6~cm continuum emission from Mart\'{\i} \et\ 
(1993); 
{\it grey thick contour} map of the CS \J{2}{1}\ integrated emission over 
the 9.8-13.6~\kms\ \vlsr\ interval; 
{\it dotted thin contour} map of the averaged CO \J{1}{0}\ blueshifted emission
over the 4.7-9.7~\kms\ \vlsr\ interval;   
{\it dashed thin contour} map of the averaged CO \J{1}{0}\ redshifted emission 
over the 12.7-17.7~\kms\ \vlsr\ interval.  
CS contours are 0.5, 1.0, 1.7, 2.4, 3.1~\jy~\kms.  CO contours are 3, 5, 8, 7, 
11 $\times$ 0.14~\jy, the noise level of the maps.  The cross indicates the 
position of the ammonia peak intensity \citep{girart94}. The solid arrow line 
shows the direction of the HH 80/81/80N flow. The half-power contour of the CS 
and CO synthesized beams ($15\farcs6\times7\farcs1$, $PA=3\arcdeg$ and 
$14\farcs9\times8\farcs0$, $PA=3\arcdeg$, respectively) are shown on the 
bottom left-hand corner, with the CS beam to the right.
The top panels show the CO \J{1}{0}\ spectra averaged over the BE and RW lobes.
\label{fmapa}
}
\end{figure}

\begin{figure} 
\epsscale{0.5}
\plotone{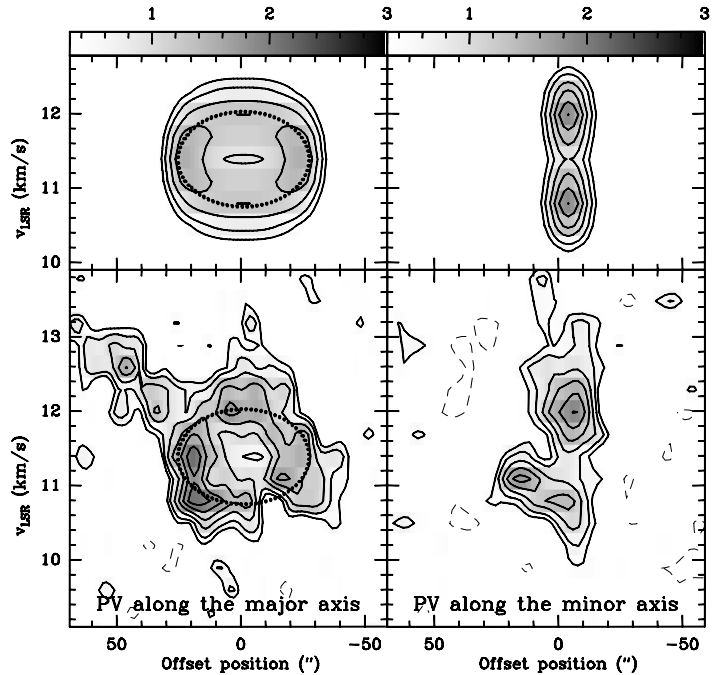} 
\figcaption[fposvel.ps]
{{\it Bottom left panel:} Position-Velocity (PV) plot of CS \J{2}{1}\ emission 
along the major axis of the dense core, $PA = 125\arcdeg$, with the position 
offset relative to $\alpha (J2000) = 18^{\rm h}19^{\rm m}18\fs13$;  
$\delta (J2000) = -20\arcdeg40'40\farcs4$.  
{\it Bottom right panel:} PV plot of CS \J{2}{1}\ emission along 
the minor axis of the dense core.
{\it Top panels:}
Synthetic emission of the PV plots along the major (right) and minor (left) 
axis of an infalling ring with an infall velocity of 0.59~\kms, no rotation, 
an inner and outer radius of 0.19 and 0.29~pc, respectively, and line width 
of 0.8~\kms.
\label{fposvel}
}
\end{figure}

There are two observational features that suggests that HH80/81/80N outflow 
may be interacting with the ambient gas surrounding the infalling main core:
First, the clear velocity gradient of the RSE arm seen in both CS and CO 
emission and its location just above \hh\ (see Fig.~\ref{fposvel}). Second, 
the BE lobe of the molecular outflow spreads out to the north at the position 
where it coincides spatially with the axis of the HH80/81/80N outflow (see 
Fig.~\ref{fmapa}: the arrow line indicates the outflow direction of 
\hh).  This scenario, where \hh\ and the dense gas are almost co-spatial, 
suggests that the infalling dense core and the RSE arm may still be affected 
chemically by the strong UV radiation of \hh, as suggested by Girart \et\ 
(1994, 1998), since this radiation is so strong that it has created 
a PDR region in its closest vicinity (Molinari \et\ 2001).  

It is unlikely that \hh\ triggered the star formation within the main core:
assuming a \hh\ propagation speed of 100--300~\kms, and using the typical age 
of a Class 0 source for the YSO embedded in the main core, $\sim 10^4$~yrs, 
\hh\ would have to be about 1 to 3~pc away from the main core when the 
collapse was initiated.  Yet, the material from the HH80/81/80N 
flow was likely present near to the CS core before \hh\ arrived to its 
proximity: a giant bow shock is observed dowstream of HH80/81 
(Heathcote \et\ 1998) and the analysis of the line emission from HH80/81/80N 
shows that the shocks arise at the interface between two fast-moving flows
(Molinari \et\ 2001). Therfore, it is still possible that the HH80/81/80N 
outflow has triggered or at least speed up the star formation.  There are 
many uncertainties in our understanding of the dynamics of collapse; these 
dynamics are determined by the balance of gravity against pressure.  Once the 
collapse is triggered, it is yet not clear how infall proceeds since thermal 
and non-thermal pressure may support the clump from collapsing, or at least 
retard an, often assumed, free-fall collapse \citep{rawlings92}.  In the case 
of \hh, the UV radiation from the shock front may play an important role in 
the dynamics of the collapse of the dense core since it will highly affect 
its cooling function.  \citet{viti99} showed that the radiation generated in 
the HH shock drives a rapid chemistry in the dense core, by lifting the icy
mantles and therefore enhancing the gas with hydrogenated species, such as
NH$_3$ and H$_2$O; the latter, for example, will undoubtedly cool the gas.  
This would `encourage' the infall.  This scenario will be investigated in
a further paper.

\acknowledgments
We would like to thank Jorge Cant\'o, Frank Shu and the anonymous referee for 
their valuable comments.  JMG is supported by NSF grant AST-99-81363 and 
by RED-2000 from the Generalitat de Catalunya.  RE is partially supported by 
DGICYT grant PB98-0670 (Spain).  SV and DAW thank PPARC for supporting their
research.

%\clearpage

%References%%%%%%%%%%%%%%%%%%%%%%%%%%%%%%

\clearpage

%Figures%%%%%%%%%%%%%%%%%%%%%%%%%%%%%%%%%

%11111111111111

%22222222222222

%Tables%%%%%%%%%%%%%%%%%%%%%%%%%%%%%%%%%%

%% TABLE 1
%\clearpage
\begin{deluxetable}{lc}
\tablecolumns{2}
\tablewidth{240pt}
\tablecaption{Physical Parameters of the \hh\ region\tablenotemark{a}
\label{fisica}}
%\tablehead{}
\startdata
\hline
\sidehead{Molecular Dense Core:} 
   $M$ (\mo)                   & $12 [X/2\times10^{-9}]^{-1}$
\\ $V_{\rm core}$ (\kms)       & 11.43      \tablenotemark{b}
\\ $\Delta V_{\rm core}$ (\kms)& 0.7        \tablenotemark{b}
\\ $R_{\rm inner}$ (pc)        & 0.19       \tablenotemark{b}
\\ $R_{\rm outer}$ (pc)        & 0.29       \tablenotemark{b}
\\ $V_{\rm rotation}$  (\kms)  & $\la 0.20$ \tablenotemark{b}
\\ $V_{\rm infall}$  (\kms)    & $0.59$     \tablenotemark{b}
\\ $\dot{P}_{\rm infall}$ (\mo\ \kms\ \yr) & $2\times10^{-5}$
\\ $\dot{M}_{\rm infall}$ (\mo\ \yr)       & $3\times10^{-5}$
\\ 
\sidehead{Molecular Outflow:} 
  $M$ (\mo)                   & 0.11
\\ $R_{\rm max}$ (pc)          & $\sim$0.4
\\ $V_{\rm max}$ (\kms)        & 10
\\ $\tau_{\rm d}$ (yr)         & $3.6\times10^4$
\\ $P$ (\mo\ \kms)             & 0.35
\\ $\dot{P}$ (\mo\ \kms\ \yr)  & $1.0\times10^{-5}$ 
\\ $L_{\rm mech}$ (\lo)        & $1.1\times10^{-3}$
\enddata
\tablenotetext{a}{Assuming optically thin emission, LTE and no
correction for inclination of the outflow}
\tablenotetext{b}{From the infalling ring model 
(see \S~\ref{cau}).}
\end{deluxetable}

\end{document}